\documentclass[journal,twocolumn]{IEEEtran}
\usepackage{amsfonts}

\usepackage{cite}

\ifCLASSINFOpdf
\else
\fi
\usepackage{mathtools}

\usepackage{graphicx}
\usepackage{amsmath,amssymb,bm}

\newtheorem{definition}{Definition}

\usepackage{caption}

\usepackage{booktabs, multirow} 
\usepackage{soul}
\usepackage[table]{xcolor} 
\usepackage{changepage,threeparttable} 
\usepackage{subcaption}

\begin{document}
%
\title{ Sparse Multi-Decoder Recursive Projection Aggregation for Reed-Muller Codes}
%
%
%


\author{Dorsa~Fathollahi, Nariman~Farsad,
        Seyyed~Ali~Hashemi, Marco~Mondelli
\thanks{D.~Fathollahi and M.~Mondelli are with the Institute of Science and Technology (IST) Austria,
Klosterneuburg, Austria (email: dorsa.fathollahi@ist.ac.at, marco.mondelli@ist.ac.at).}
\thanks{N.~Farsad is with the Department of Computer Science, Ryerson University, Toronto, Canada (email: nfarsad@ryerson.ca).}
\thanks{ S.~A.~Hashemi is with the Department of Electrical Engineering, Stanford University, Stanford, CA 94305, USA (email: ahashemi@stanford.edu).}
}

\maketitle

\begin{abstract}
Reed-Muller (RM) codes are one of the oldest families of codes. Recently, a recursive projection aggregation (RPA) decoder has been proposed, which achieves a performance that is close to the maximum likelihood decoder for short-length RM codes. One of its main drawbacks, however, is the large amount of computations needed. In this paper, we devise a new algorithm to lower the computational budget while keeping a performance close to that of the RPA decoder. The proposed approach consists of multiple sparse RPAs that are generated by performing only a selection of projections in each sparsified decoder. In the end, a cyclic redundancy check (CRC) is used to decide between output codewords. Simulation results show that our proposed approach reduces the RPA decoder's computations up to $80\%$ with negligible performance loss. 

\end{abstract}


%
\IEEEpeerreviewmaketitle

\section{Introduction}
\label{sec:intro}

	Reed-Muller (RM) codes were introduced by Muller \cite{6499441}  in 1954. Shortly after, Reed proposed a majority logic decoding algorithm \cite{1057465} correcting errors up to half of its minimum distance. Since then, many approaches for decoding RM codes have been investigated. An overcomplete minimum weight parity check matrix is used in \cite{1057216} to exploit the redundant code constraints; recursive list decoding in \cite{1291729, 1603792, 1603764} achieves a performance close to maximum likelihood (ML) decoding with list size at-most $1024$ for short block lengths; and the Sidel'nikov-Pershakov algorithm \cite{zbMATH00223749} decodes second-order RM codes of length $ \leq 1024$  by exploiting derivatives of the received codeword and majority voting. Additionally, Sakkour's  \cite{1531882}  variant of \cite{zbMATH00223749}  simplifies the majority voting leading to achieving a smaller decoding error probability.

	Recently, RM codes have received a great deal of attention. One reason for this surge of interest is their close connection to polar codes, a family of error-correcting codes that provably achieves capacity for any binary-input memoryless symmetric channel \cite{5075875}. This connection was already mentioned in the seminal paper \cite{5075875}, and it was exploited to design a family of interpolating codes with improved performance at practical block lengths in \cite{mondelli2014polar}. Furthermore, it has been shown that, under  ML decoding, RM codes achieve capacity on erasure channels \cite{7862912}. Achieving similar results for a more general class of channels is a long-standing conjecture.

In general, ML decoding is not computationally efficient. This has spurred research on finding an efficient algorithm. Many approaches taking advantage of different aspects of RM codes have been exploited recently. In \cite{8437637}, minimum-weight parity checks are employed taking advantage of the large automorphism group of RM codes. Berlekamp-Welch type algorithms on random errors and erasures are considered and analyzed in  \cite{7858796, 10.5555/3381089.3381171}. Successive cancellation (SC) decoding \cite{5075875} and SC list (SCL) decoding \cite{7055304}, initially proposed for polar codes, are also applicable to RM codes as they share a similar construction. An algorithm based on successive factor graph permutations is presented in \cite{hashemi2018decoding}. For a thorough review on RM codes, we refer the reader to \cite{9123985}.

Most recently, a recursive projection aggregation (RPA) \cite{RPAAY} decoder has been proposed. The RPA decoder uses projections of the codeword on its index space cosets in order to obtain valid RM codewords of smaller length. Then, the smaller codewords are recursively decoded and aggregated. RPA decoding performs close to ML decoding up-to length $1024$ for second-order RM codes and also benefits from a parallel implementation. A recursive puncturing aggregation (RXA) decoder is presented in \cite{9174087}, and it is a modification of RPA decoding using puncturing instead of projections, built for high-rate codes.

In this paper, we focus on the RPA decoding algorithm. The near ML decoding performance benefits of the RPA decoder comes with the need for a large computational budget. We propose an optimized version of RPA decoding, namely a sparse RPA with multiple decoders (SRPA). The new method chooses a subset of recursions for each decoder at random. When combined, these smaller decoders can perform close to RPA decoding while requiring a significantly smaller computational budget.

We compare the proposed SRPA method with RPA decoding on code lengths  $\leq 512$. The results show that by using only two sparse decoders, one can achieve almost the same performance at $20\%$ of the computational budget of RPA decoding. Also, the performance of SRPA is compared with SCL decoding while fixing the computational budget. The results indicate that for second-order RM codes, the performance of the proposed SRPA is more stable, staying relatively close to the performance of ML decoding as the block length increases.

	The rest of the paper is organized as follows. In Section~\ref{sec:def}, we give the basic definition of RM codes and the description of the RPA decoder. In Section~\ref{sec:problem}, we discuss the problem formulation and the proposed method. In Section~\ref{sec:sim}, we present the simulation results. Finally, in Section~\ref{sec:conc}, we draw the main conclusions of the paper.

\section{Basic Definitions}
\label{sec:def}

\subsection{Reed-Muller Codes}

    There are many approaches to describing RM codes. Here we define them by using low-degree polynomials following the notations of \cite{RPAAY}.
    
	Consider $\mathbb{F}_2[Z_1 , Z_2 , \ldots , Z_m] $ as the polynomial ring of $m$ variables. Since $Z^2 = Z$  in $\mathbb{F}_2 $, then the following set of monomials with size $2^m$ forms a basis for $\mathbb{F}_2[Z_1 , Z_2 , \ldots , 	 Z_m] $:

	\begin{equation}
	\{\prod_{i\in A} Z_i | A \subseteq [m]  \}, 
	\end{equation}
	where $[m] \coloneqq \{1,2,\ldots, m\}$  and $\prod_{i\in \emptyset} Z_i \coloneqq 1$.
	
	For each subset $A\subseteq [m] $ let $\bm{v}_m(A)$ be a vector of length $2^m$,  indexed by $\bm{z}= (z_1 , z_2 , \ldots , z_m ) \in \{ 0, 1 \}^ m $, such that $\bm{v}_m(A,\bm{z})$, the $\bm{z}$-th component of $v_m$, is defined  as
	
	\begin{equation}
	\bm{v}_m(A,\bm{z})= \prod_{i \in A } z_i.
	\end{equation}

	\begin{definition}[Reed-Muller codes]
		The $r$-th order \textit{Reed-Muller} code $\mathcal{R}\mathcal{M} (m,r) $ is defined as
		\begin{equation}
		\begin{split}
		\mathcal{R}\mathcal{M} (m,r) \coloneqq \{ & \sum_{A\subseteq [m] , |A| \leq r }  u(A)\bm{v}_m(A): \\ & u(A)\in \{0,1\},   \forall A\subseteq [m] , |A|\leq r   \}.
		\end{split}
		\end{equation}
			\end{definition}
This code has dimension $\sum_{i = 0}^{r} \binom{m}{i}$ and length  $n=2^m $.

\subsection{Recursive Projection Aggregation Decoder}

For a binary-input memoryless channel $W : \{0,1\} \rightarrow \mathcal{W} $ the log-likelihood ratio (LLR) of the channel output is defined as

\begin{equation}
    LLR(x) \coloneqq \ln \left( \frac{W(x|0)}{W(x|1)} \right).
\end{equation}

Given $L$ as a vector corresponding to the LLR values of the channel output, the RPA decoder obtains codewords of a shorter length by projecting $L$ onto the cosets of the index subspaces. The shorter codewords are decoded recursively. Later, the results are aggregated via a voting procedure to estimate the original codeword.

Let $\mathbb{E} \coloneqq \mathbb{F}_2^m$ and let each codeword correspond to an $m$-variate polynomial of degree at-most $r$. Then, the coordinates of the codeword $c\in \mathcal{R}\mathcal{M} (m, r) $ are indexed by the binary vectors $\bm{z}\in \mathbb{E}$, i.e., $c = (c(\bm{z}) , \bm{z} \in \mathbb{E}) $.
	
	    \begin{figure}

    	\centering
    	\includegraphics[width=\linewidth]{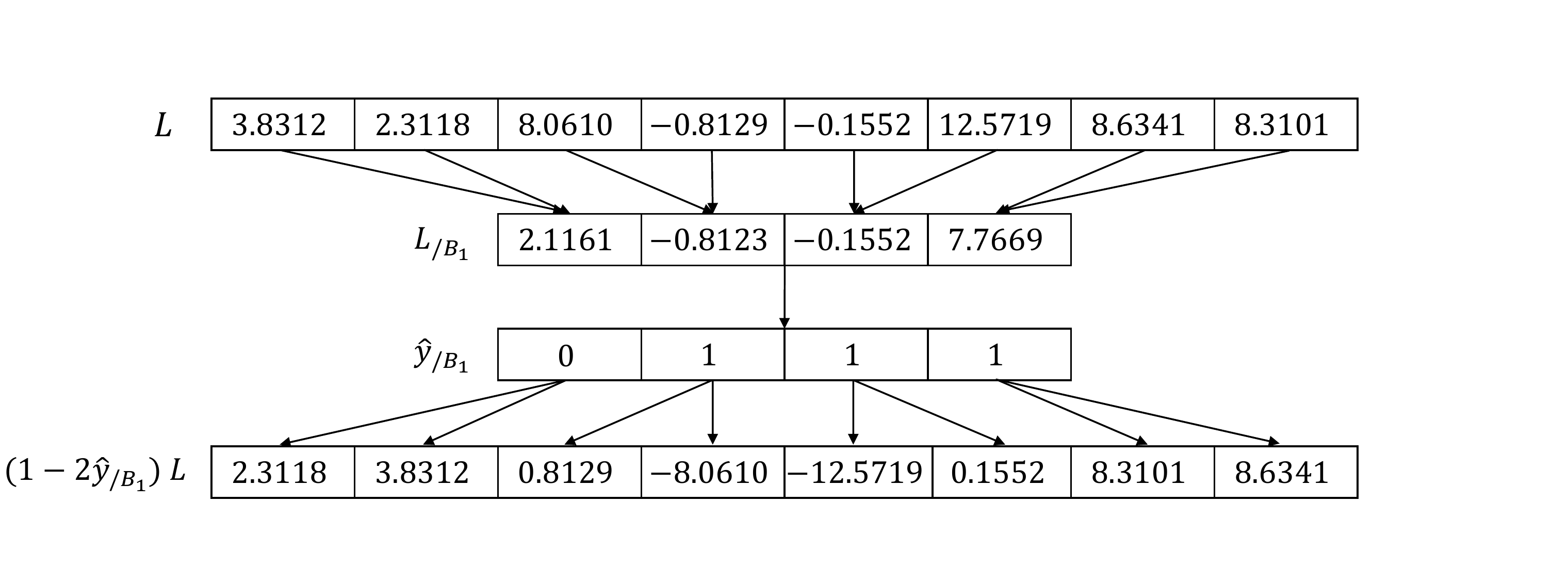}
    	\caption{RPA single projection and aggregation example on $\mathcal{R}\mathcal{M}(3,2)$ and cosets of $\mathbb{B}_1$. }
         \label{fig:ex-proj-agg} 
    \end{figure}

	\begin{definition}[Projection]
		 Let $\mathbb{B}$ be an $s$-dimensional ($s\leq r $) subspace of $\mathbb{E}$ and $\mathbb{E} / \mathbb{B} $ be the set of all cosets of $\mathbb{B}$ in $\mathbb{E}$; in other words, $ \forall T \in \mathbb{E}/\mathbb{B}$ , $T = \bm{z} + \mathbb{B}$ for some $z \in \mathbb{E}$. The projection of the codeword $y$ on the cosets of $\mathbb{B}$ for the binary symmetric channel (BSC) is defined as
		
		\begin{equation}\label{eq:projBSC}
		y_{/\mathbb{B}} = Proj(y,\mathbb{B}) \coloneqq (	y_{/\mathbb{B}} (T) , T\in \mathbb{E} / \mathbb{B}),
		\end{equation}
		where
		\begin{equation}
		y_{/\mathbb{B}}(T) \coloneqq \bigoplus_{\bm{z} \in T} \  y(\bm{z}).
		\end{equation}
		For general binary-input channels having vector $L$ as LLR values of the noisy codeword, the definition above is modified into
		\begin{equation}\label{eq:proj} 
		\begin{split}
		L_{/ \mathbb{B}  }(T)= &\ln \left(\exp \left(\sum_{\bm{z} \in T} L(\bm{z} )\right)+ 1\right) \\ & - \ln \left(\sum_{\bm{z}\in T} \exp\left(L(\bm{z})\right)\right).
		\end{split}
		\end{equation}

	\end{definition}
	
	Note that, from Lemma 1 in \cite{RPAAY}, we have that if $y\in \mathcal{R}\mathcal{M} (m,r) $, then $	y_{/\mathbb{B}} \in \mathcal{R}\mathcal{M} (m-s,r-s)$.

	The RPA algorithm is obtained by setting $s=1$. In this case, at level $i$ in the recursion, each codeword will make $2^{n-i} -1 $ branches. After obtaining the estimated codewords for each branch, the results are aggregated using the voting algorithm below.

	\begin{definition}[Aggregation/Voting Algorithm] Let $L$ be the output LLRs,  $\hat{y}_{\mathbb{B}_i} $ be the decoded codeword corresponding to the cosets of the one dimensional subspaces $\mathbb{B}_i = \{0, \bm{z}_i\}  ,  \bm{z}_i\in \mathbb{F}_2^m $, and $\hat{L}$ be the final estimation. For each bit index $\bm{z}\in \{0,1\}^m $, the estimation is computed as  
		\begin{equation}\label{eq:agg1} 
	\hat{L}(\bm{z}) = \frac{1}{n} \sum_{i = 1 }^{n-1} \left(\left(1 - 2 \hat{y}_{ / \mathbb{B}_i}\left([\bm{z}+\mathbb{B}_{i}]\right)\right)L\left(\bm{z}\oplus \bm{z}_i\right)\right),
	\end{equation}
where $[\bm{z} + \mathbb{B}_i]$ is defined as the coset containing $\bm{z}$. 
	\end{definition}

One way to look at (\ref{eq:agg1}) is that for a one dimensional subspace $\mathbb{B}_i$ and index $\bm{z}\in \mathbb{F}_2^m \setminus 0$, if $\bm{z}^\prime = \bm{z}\oplus \bm{z}_i $, the decoding result of the corresponding coset, $(1 - 2 \hat{y}_{ / \mathbb{B}_i}\left([\bm{z}+\mathbb{B}_{i}]\right))$, is an estimation of $c(\bm{z}) \oplus c(\bm{z}^\prime) $. If $c(\bm{z})\oplus c(\bm{z}^\prime)$ is more likely to be $1$, then their sign of LLR values should be different. On the other hand, if the sum is more likely to be $0$, then their LLR values should also have the same sign. Fig.~\ref{fig:ex-proj-agg} shows an example of this procedure on $\mathcal{R}\mathcal{M}(3,2)$ and cosets of $\mathbb{B}_1$. 
    
 	The algorithm starts from $\mathcal{R}\mathcal{M}(m,r)$ and recursively decodes $\mathcal{R}\mathcal{M}(m-1,r-1 )$ until reaching $\mathcal{R}\mathcal{M}(m-r+1 , 1 )$. For first-order RM codes, the ML decoder has complexity $O(n\log n)$ and can be implemented using the Fast Hadamard Transform (FHT) \cite{1057189}. According to \textit{Algorithm~3} in \cite{RPAAY},  the RPA decoding algorithm is iterative, in the sense that it  repeats the procedure above at most $\frac{m}{2}$ times until it reaches a fixed point where $|L (\bm{z}) - \hat{L} (\bm{z}) | \leq \epsilon$, $\forall \bm{z}\in \mathbb{F}_2^m$, for some small assigned constant $\epsilon$.
 	

\section{Sparse Recursive Projection Aggregation}
\label{sec:problem} 
	Although the RPA decoding algorithm performs well on RM codes with respect to the block error rate (BLER), it requires significant computing power. Given an $r$-th order RM code of block length $n$, the algorithm runs in $O(n^r \log n )$ time for a sequential implementation and in $O(n^2)$ time for a parallel implementation using $O(n^r)$ processors. That is because at the $i$-th level of recursion, the RPA decoder uses $n2^{-i} -1 $ recursive calls to vote for each sub-code. As we will see in Section~\ref{subsec:pruning}, having this amount of recursive calls is redundant for obtaining a reliable vote. Our goal is to achieve the same error probability to that of the RPA decoder using lower computational power. We will achieve this goal by reducing the number of recursive calls.

	\subsection{Random Selection of Recursions}\label{subsec:pruning}
	In order to reduce the amount of computation required, several recursions are chosen at random to be kept and the rest are removed to generate a sparse decoder as defined below.
    \begin{definition}\label{def:SRPA}
    A sparse decoder $\mathcal{D}_S$ (SRPA) that has at most $t$ iterations and $q$ recursions/subspaces per iteration, performs only a subset $S= \{ s_{1,1} , \ldots, s_{t,q} \}$ of possible projections, where $s_{i,j}$ is the $j$-th chosen subspace in the $i$-th iteration that the codeword is projected on.
    \end{definition}
    
    
    Definition~\ref{def:SRPA} suggests that the aggregation formula in  (\ref{eq:agg1})  for each  $\bm{z} \in \{ 0, 1 \}^m$ is modified to
    \begin{equation}\label{eq:agg2}
    	\hat{L}(\bm{z}) = \frac{1}{q} \sum_{j \in S_i }  \left(\left(1 - 2 \hat{y}_{ / \mathbb{B}_j}\left([\bm{z}+\mathbb{B}_{j}]\right)\right)L\left(\bm{z}\oplus \bm{z}_j\right)\right),
    \end{equation}
    where $S_i = \{ s_{i,1} , s_{i,2} , \ldots, s_{i,q}  \}$ is the subset of $S$ corresponding to the $i$-th iteration of SRPA decoding.
  	
    \begin{figure}

    	\centering
    	\includegraphics[width=2.3in]{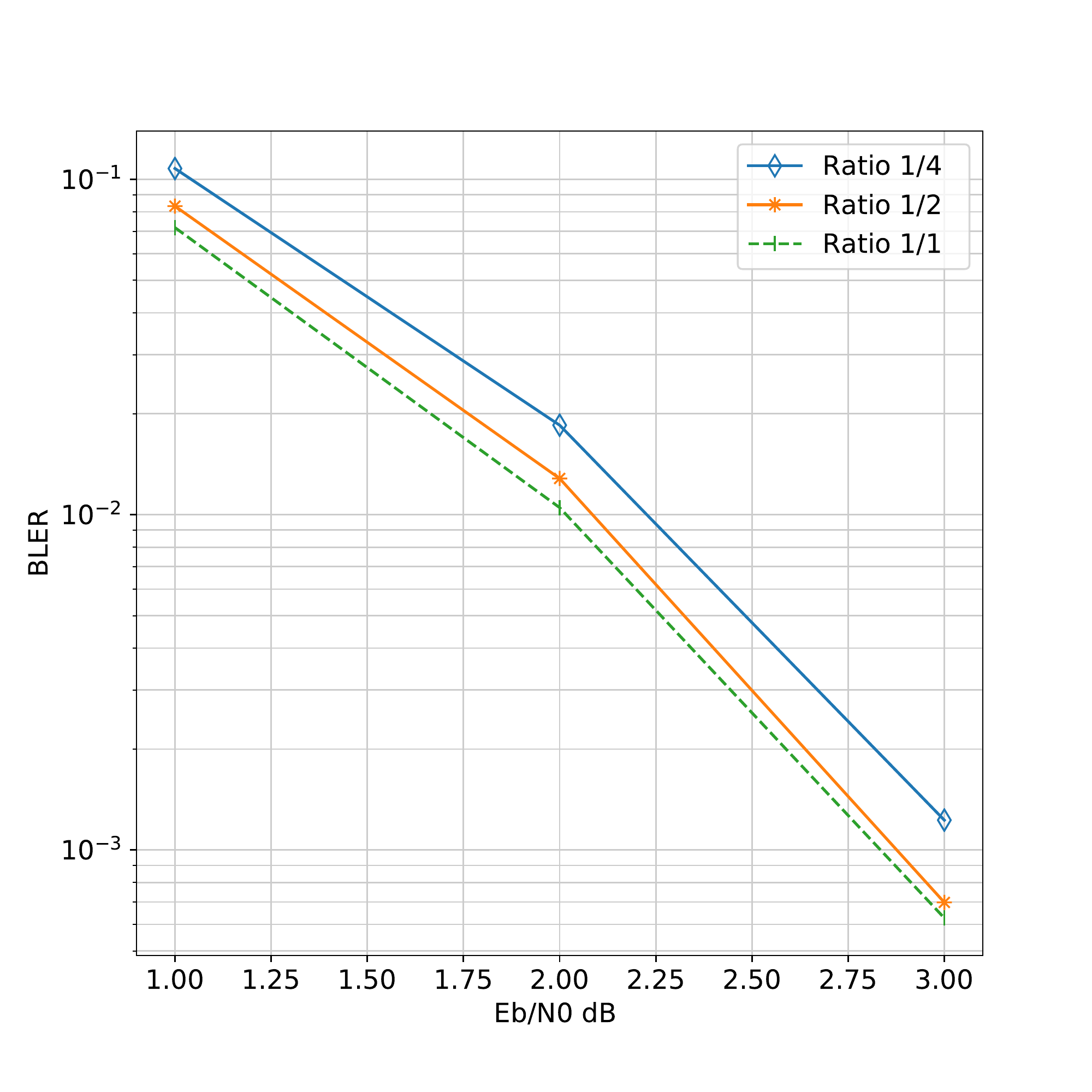}

    	\caption{Different ratios of remaining recursions on $\mathcal{R}\mathcal{M} (7,2) $. By keeping half of the projections, we lose only $0.1$ dB in performance; however, by keeping only $1/4$ of the projections, the performance degrades by $0.25$ dB. }
         \label{fig:1} 
    \end{figure}

    The simulation results for $\mathcal{R}\mathcal{M}(7,2)$ in Fig.~\ref{fig:1} show that removing up to half of the recursions causes a negligible loss in performance, which confirms the high redundancy of the RPA decoding algorithm. However, if only $1/4$ of the projections are kept, the error probability starts to degrade. The reason for this degradation is that each projection estimates certain codewords more reliably and by removing a large fraction of recursions, it is more likely to miss the suitable projections.
    

    \subsection{Choice of Multiple Decoders}
    
    The loss of performance that is caused by the pruning method in SRPA can be compensated by using multiple SRPA decoders. The idea is that the low-complexity decoders work well for a subset of errors. Choosing multiple low-complexity decoders at random will improve the error probability of SRPA while having lower complexity than RPA decoding.

    Let us select $k$ sparse decoders $\mathcal{D}_{S_1} ,\mathcal{D}_{S_2} , \ldots , \mathcal{D}_{S_k} $ at random. We call this set of $k$ SRPA decoders the $k$-SRPA decoder. The input LLR $L$ is individually decoded by each of the $k$ SRPA decoders resulting in estimations $\hat{L}_1,\hat{L}_2 ,\ldots,\hat{L}_k $. The final estimation is then chosen via one of the following methods: 
    \subsubsection{Most likely codeword} 
    The most likely codeword is chosen by computing $\operatorname*{argmax}_i \langle \hat{L}_i , L \rangle $, where $\langle\cdot,\cdot \rangle$ denotes the scalar product.
    \subsubsection{CRC verification}
    The final codeword is chosen by first checking the CRC condition and then choosing the most likely codeword among the codewords satisfying the CRC condition.
    
    Note that lower levels of recursions can only use the first approach in this setting, as projections do not allow the concatenation of CRC at these levels.

\section{Simulation Results} \label{sec:sim} 
    
   The proposed SRPA is tested on RM codes of up to third order, with block lengths $128$, $256$, and $512$. The performance of the proposed method is compared with the original RPA decoding algorithm. In the implementation of SRPA, the $3$-bit CRC (\textit{CRC-3-GSM}: $x^3 +x + 1 $) is used to select the final codeword.
    


\begin{figure*}[tbp]
\centering
\subfloat[$\mathcal{R} \mathcal{M} ( 7 , 2 ) $]{\includegraphics[width=2.3in]{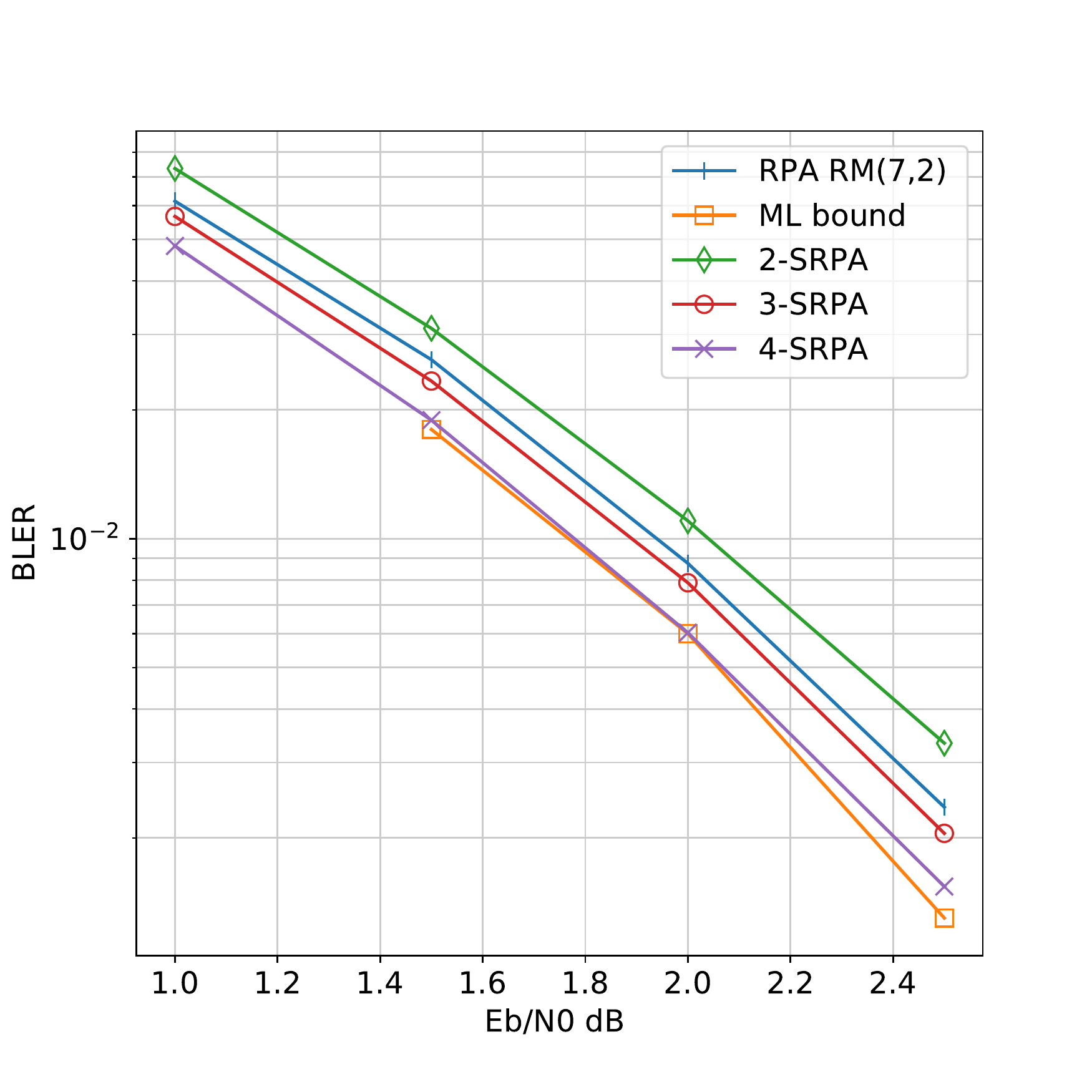}} 
\subfloat[$\mathcal{R} \mathcal{M} ( 8 , 2 ) $]{\includegraphics[width=2.3in]{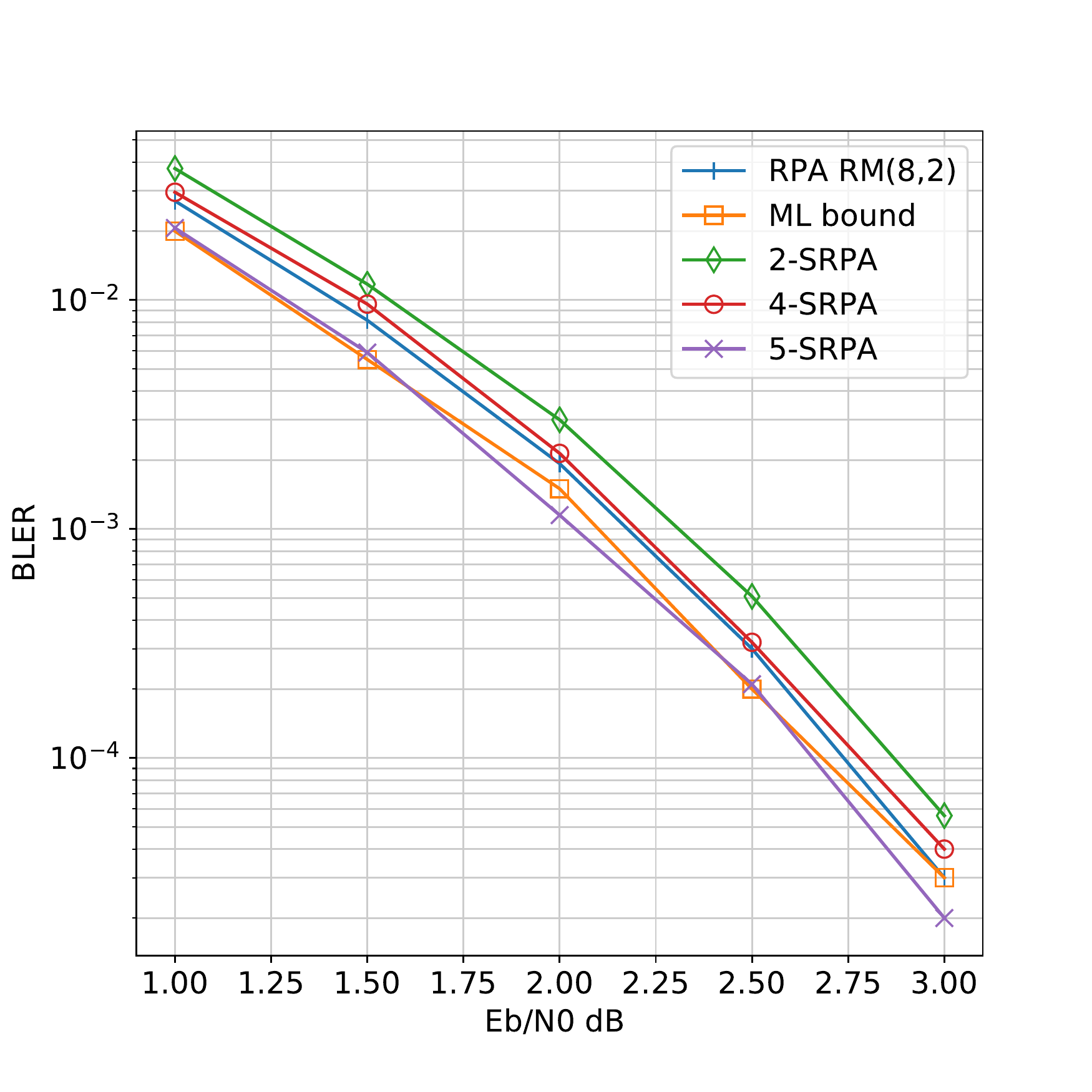}}
\subfloat[$\mathcal{R} \mathcal{M} ( 9 , 2 ) $]{\includegraphics[width=2.3in]{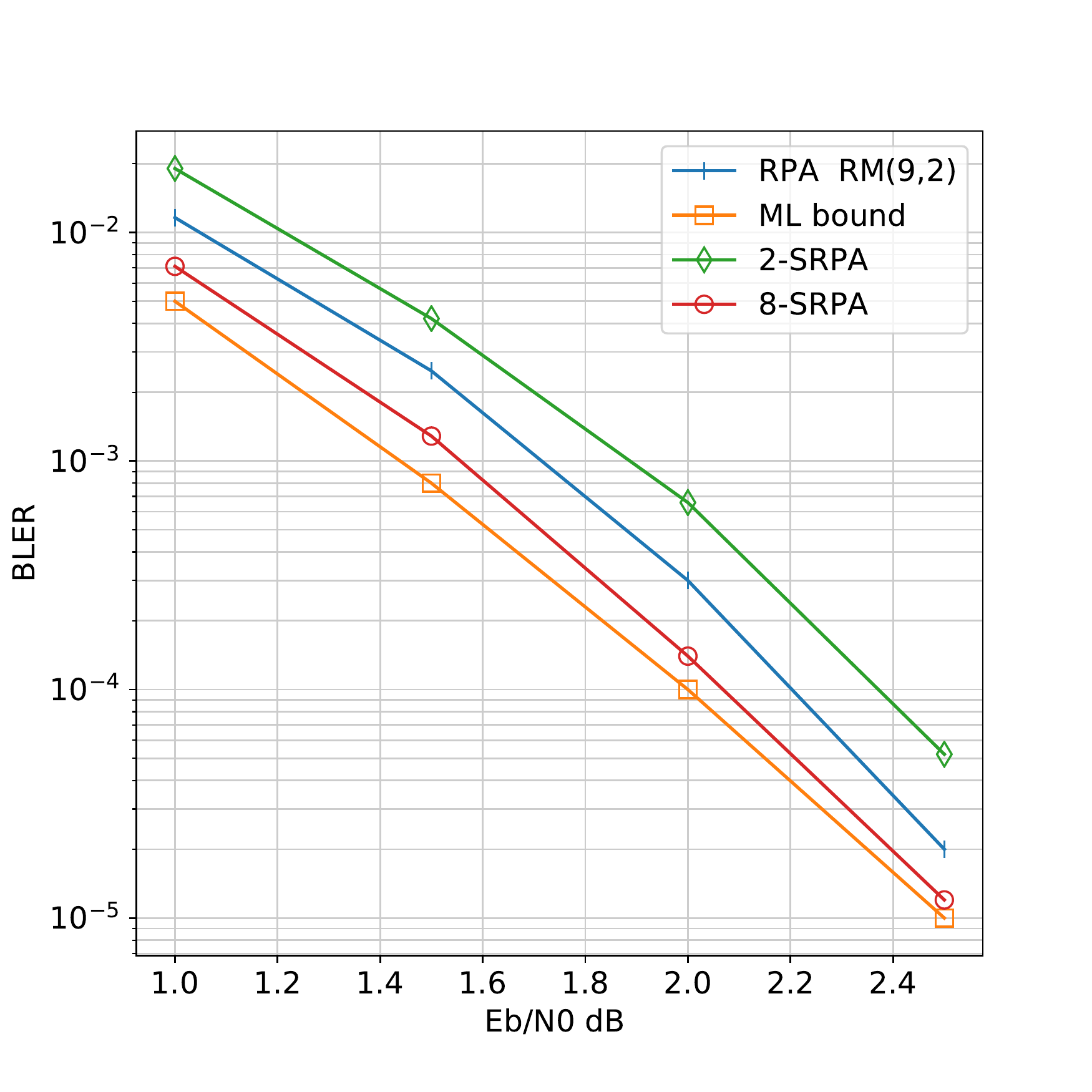}}\\
\subfloat[$\mathcal{R} \mathcal{M} ( 7 , 3 ) $]{\includegraphics[width=2.3in]{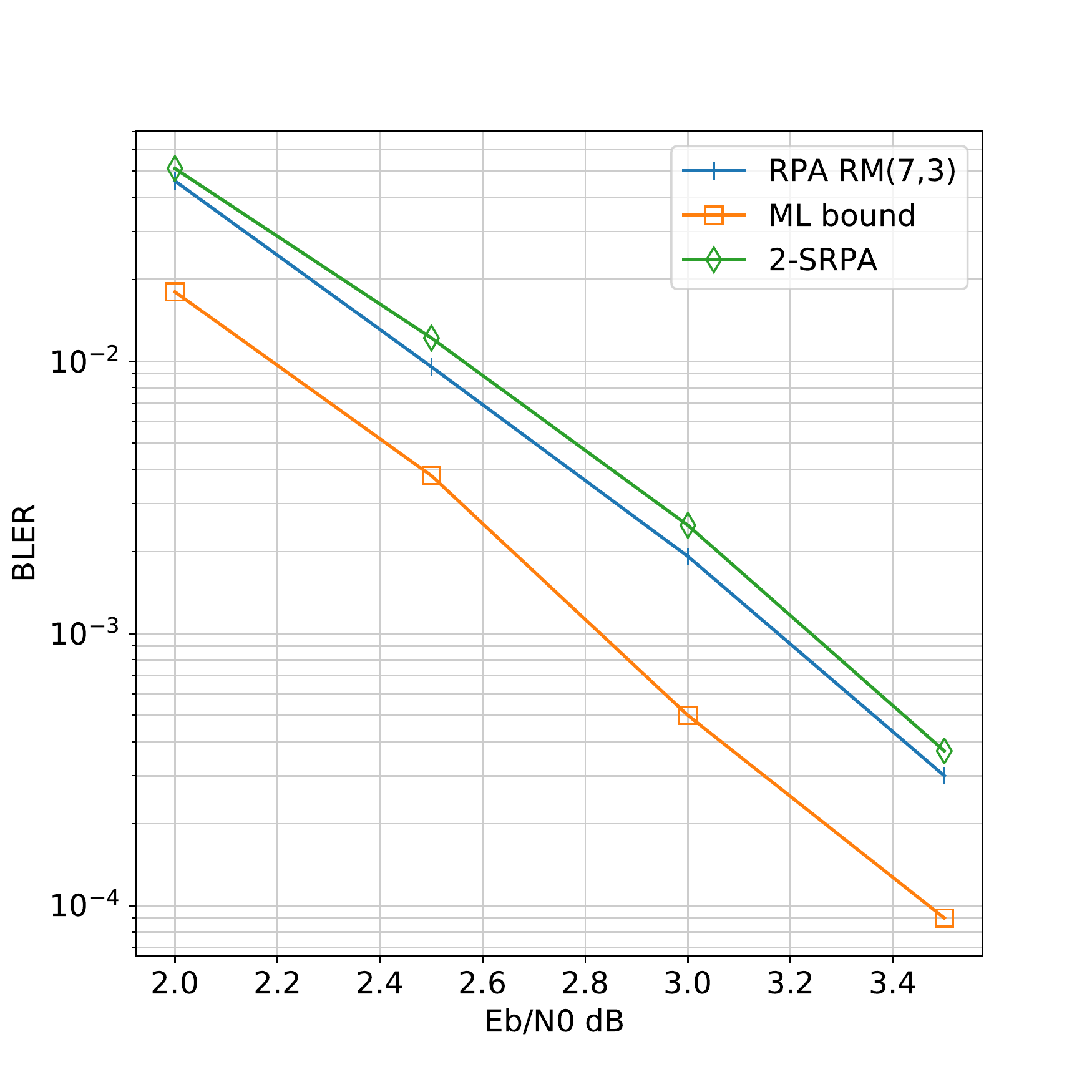}}  
\subfloat[$\mathcal{R} \mathcal{M} ( 8, 3 ) $]{\includegraphics[width=2.3in]{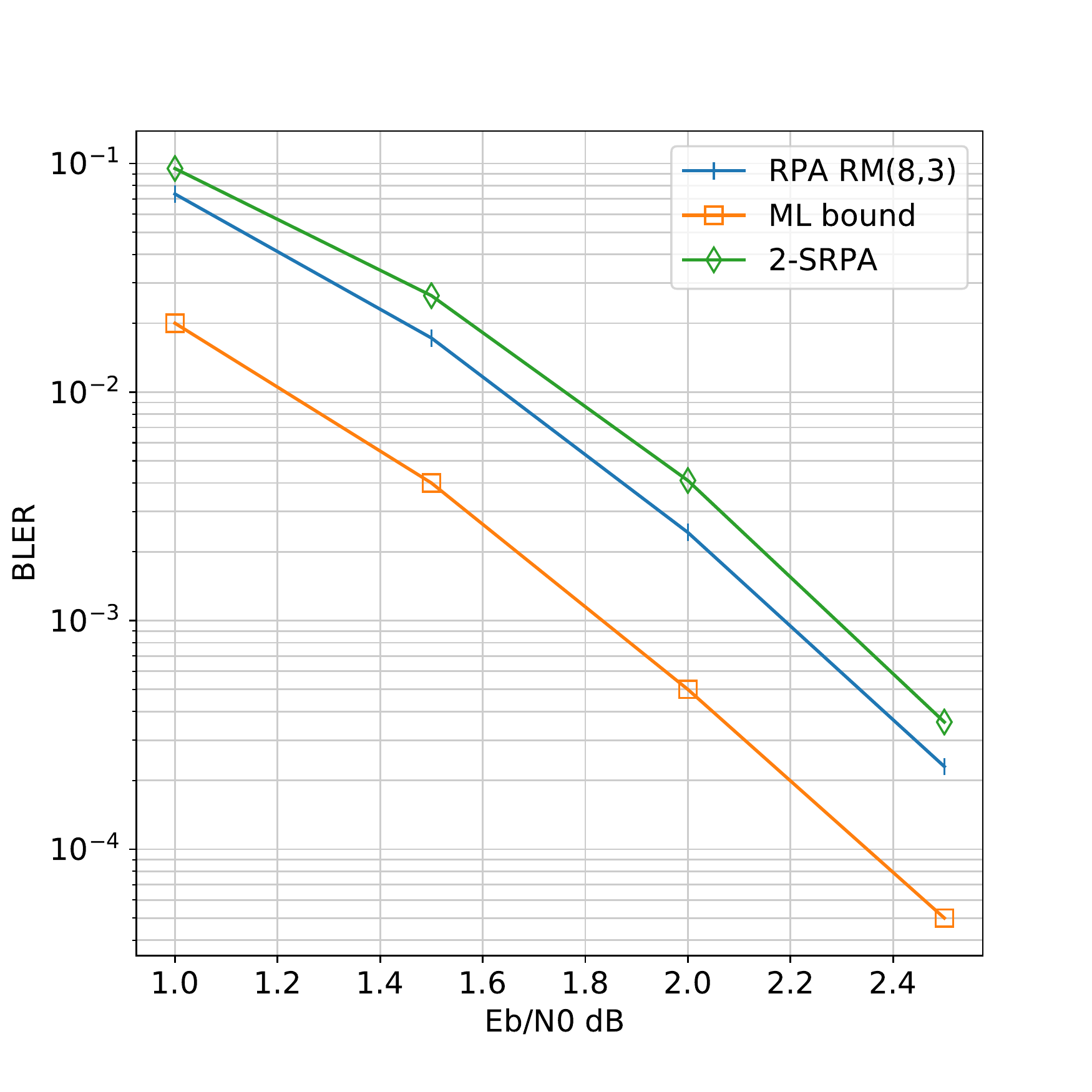}} 
\caption{ Performance comparison between $k$-SRPA decoding and RPA decoding. $2$-SRPA decoding has performance close to RPA decoding while saving up-to $87\%$ of the computational budget. For second-order RM codes, having at most $8$ SRPAs, we can achieve a performance better than RPA decoding or even approach the ML bound.}
\label{fig:2}
\end{figure*}

Note that each projection and aggregation has complexity $\Theta(n)$, and the decoding of first-order RM codes has complexity $\Theta(n\log n)$ since it is equivalent to performing an FHT. These operations are exactly the same in both RPA decoding and SRPA decoding. The difference lies in the number of operations performed, and we elaborate on this difference for RM codes of second and third order below.

In order to decode $\mathcal{R}\mathcal{M} (m,r)$, we denote by $q_i$, $t_i$ and $d_i$ the number of projections, iterations, and decoders used by SRPA for the sub-code  $\mathcal{R}\mathcal{M}(m-i,r-i)$. Then, in order to decode $\mathcal{R}\mathcal{M} (m,2)$, the total number of projections, aggregations, and FHTs is $d_0t_0q_0$. Furthermore, in order to decode $\mathcal{R}\mathcal{M} (m,3)$, we need $d_0t_0q_0 +d_0t_0q_0d_1t_1q_1$ projections, $d_0t_0q_0 +d_0t_0q_0d_1t_1q_1$ aggregations, and $d_0t_0q_0d_1t_1q_1$ FHTs. Recall that the complexity of the FHT has an extra $\log n$ factor with respect to projections and aggregation. Note also that -- in our experimental setting -- $d_0t_0q_0$ is much smaller than $d_0t_0q_0d_1t_1q_1$. Thus, in order to compare the computational budget of RPA and SRPA, it suffices to count the number of FHTs executed by the two algorithms.

    In the implementations in this paper, we use two randomly-selected decoders. For second-order RM codes, each of the decoders keep $\frac{1}{8}$ of the recursions chosen uniformly at random. Thus, SRPA roughly leads to a $4\times$ speed-up with respect to the original RPA decoding. For third order RM codes, $\frac{1}{8}$ of the first level recursions are chosen. When the recursions reach $\mathcal{R}\mathcal{M} (m-1,2) $, four decoders are used instead of two to avoid an increase in the error rate. This leads to a factor of $\frac{1}{8} \times \frac{1}{2} $ reduction in the number of FHTs. Thus, SRPA roughly leads to an $8\times$ speed-up.

    Fig.~\ref{fig:2} shows the BLER of different decoders as a function of the SNR ($E_b/N_0$). It is visible that the $2$-SRPA decoder has close performance to that of RPA decoder. In addition, by adding more decoders to the $2$-SRPA decoder for second-order RM codes, one can gain considerable improvement in the error probability and approach the performance of ML decoding.
\begin{table*}[t]
\caption{Average number of FHTs needed for RPA decoding at different $E_b/N_0$ levels, compared with $2$-SRPA decoder. The column \emph{Full Rounds} reports the number of FHTs required in RPA decoding if no early stopping occurs.}
\centering
\begin{tabular}{ccccccccccc}
\toprule
\multirow{4}{*}{Code} & \multicolumn{8}{c}{RPA} & \multirow{4}{*}{$2$-SRPA} & \multirow{4}{*}{Savings}
\\
\cmidrule(lr){2-9}
& \multicolumn{7}{c}{$E_b/N_0$ [dB]} & \multirow{2.4}{*}{Full Rounds} & &
\\
\cmidrule(lr){2-8}
& $0$       & $1$       & $1.5$     & $2$       & $2.5$     & $3$       & $3.5$     & & & \\
\midrule
$\mathcal{R}\mathcal{M}(7,2)$  &     $-$    & $366$ & $352$ & $334$ & $310$ & $288$ & $270$ & $381$        & $96$         & $75\%$ \\
$\mathcal{R}\mathcal{M}(7,3)$  &    $-$     &   $-$      &      $-$   & $38927$ & $33671$ & $29276$ & $26325$   & $73728$      & $13824$      & $81\%$ \\
$\mathcal{R}\mathcal{M}(8,2)$ & $848$ & $780$ & $755$ & $725$ & $680$ & $623$ &     $-$    & $1020$       & $256$        & $75\%$    \\
$\mathcal{R}\mathcal{M}(8,3)$ &    $-$     & $231262$  & $198009$  & $168866$  & $143056$  &    $-$     &   $-$      & $388620$     & $49152$      & $87\%$   \\
$\mathcal{R}\mathcal{M}(9,2)$ &     $-$    & $1580$ & $1543$ & $1522$ & $1484$ &    $-$     &   $-$      & $2044$       & $512$        & $75\%$   \\
\bottomrule
\end{tabular}
\label{tab:1}
\end{table*}

\begin{figure*}[tbp]
\centering
\subfloat[$\mathcal{R} \mathcal{M} ( 7 , 2 ) $]{\includegraphics[width=2.3in]{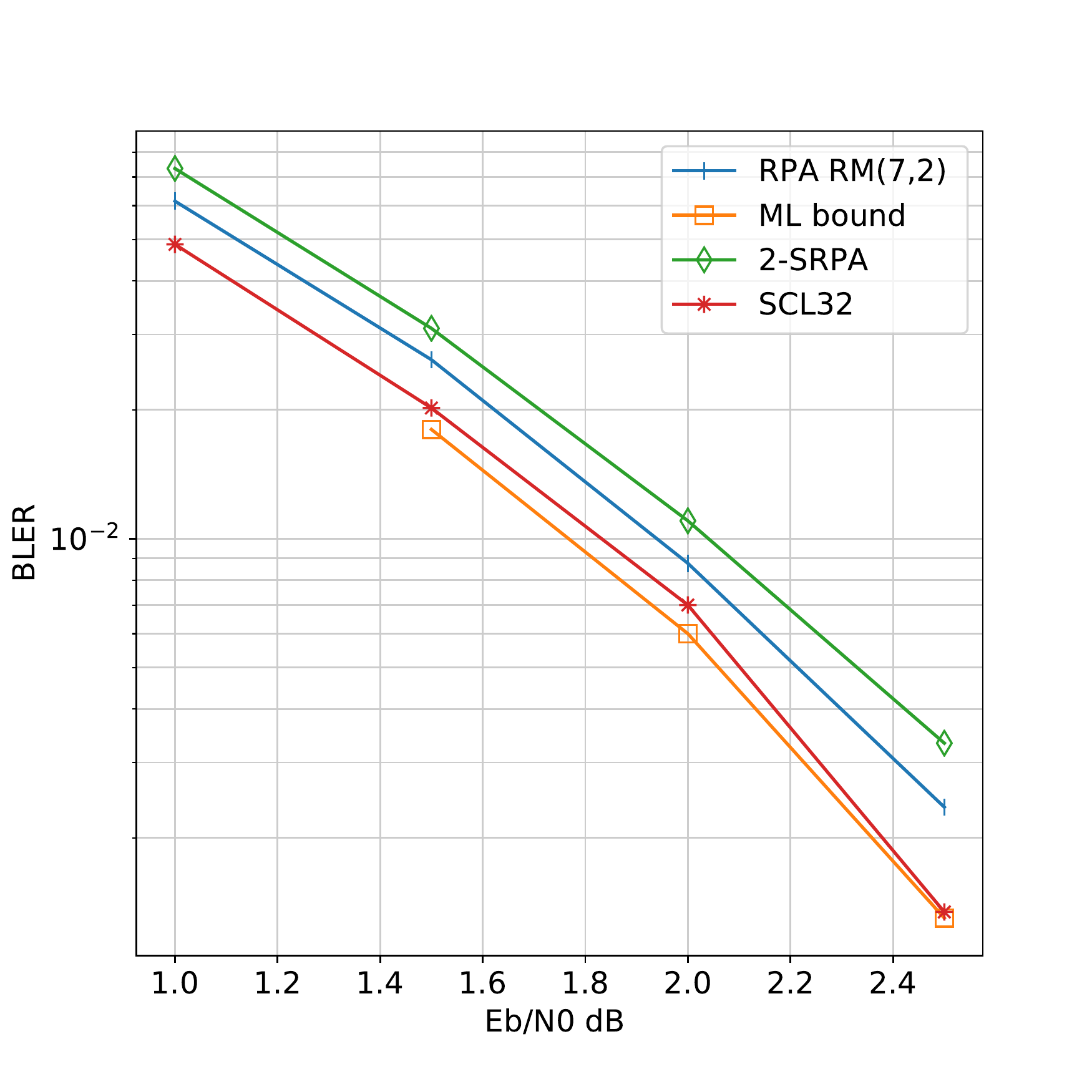}} 
\subfloat[$\mathcal{R} \mathcal{M} ( 8 , 2 ) $]{\includegraphics[width=2.3in]{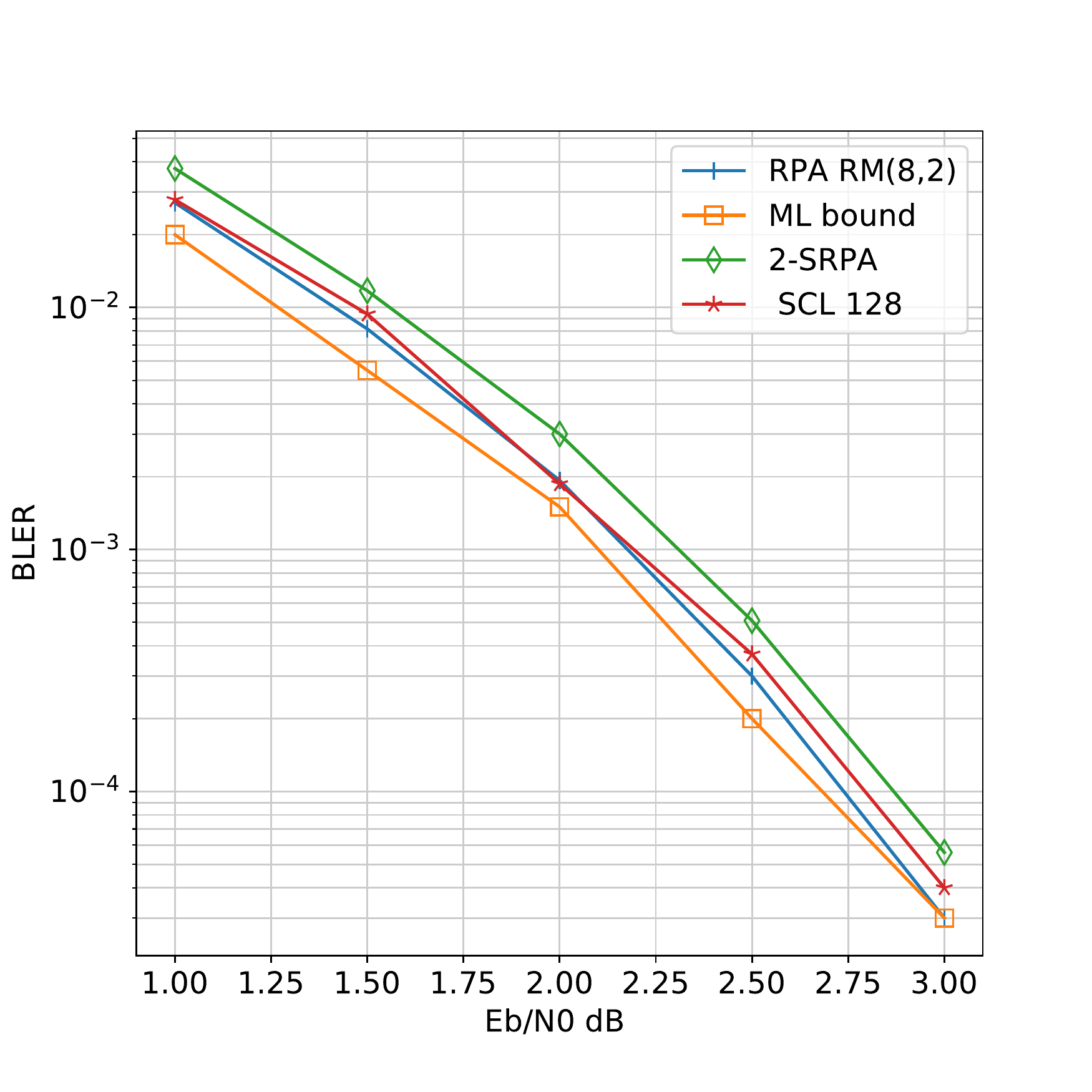}}
\subfloat[$\mathcal{R} \mathcal{M} ( 9 , 2 ) $]{\includegraphics[width=2.3in]{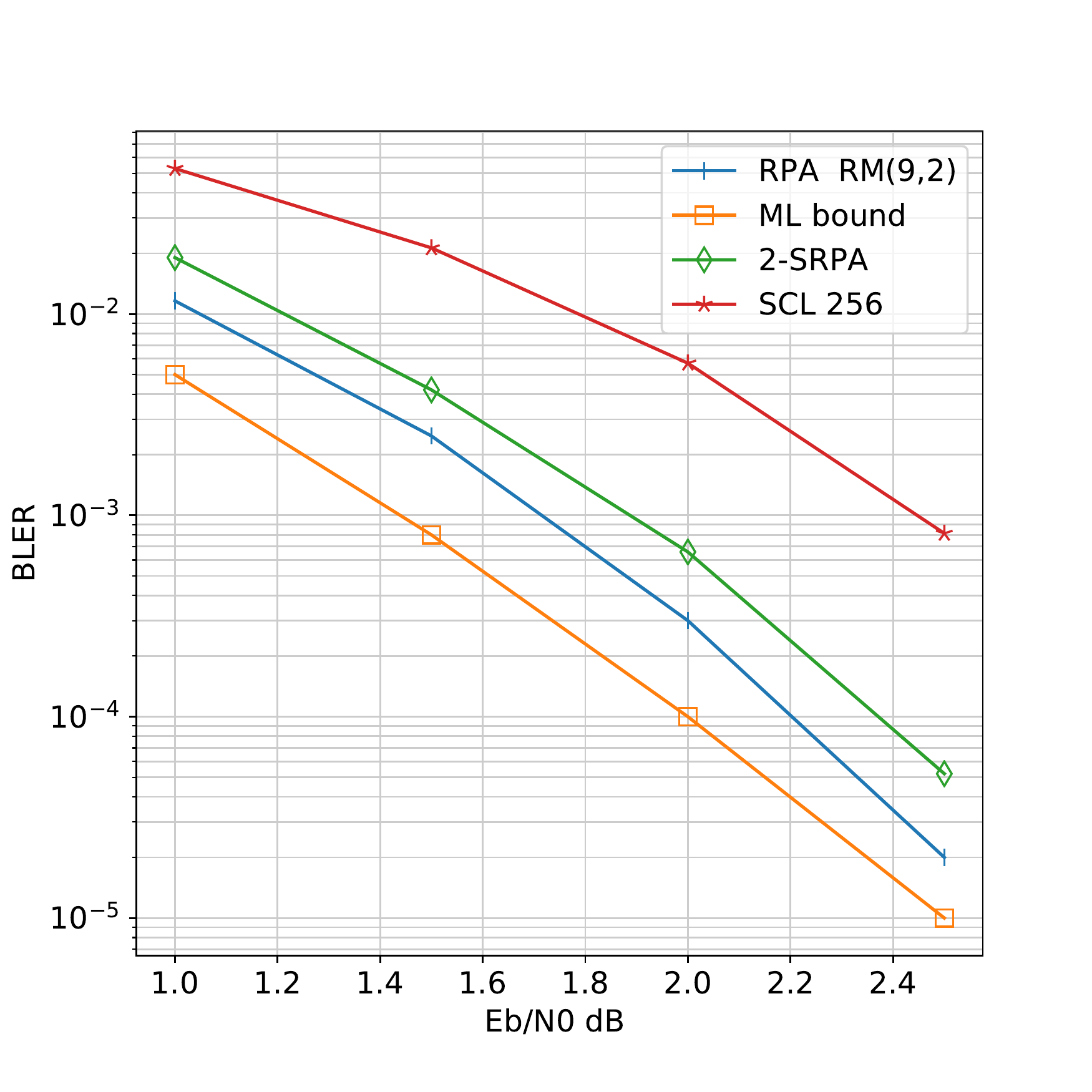}}\\
\subfloat[$\mathcal{R} \mathcal{M} ( 7 , 3 ) $]{\includegraphics[width=2.3in]{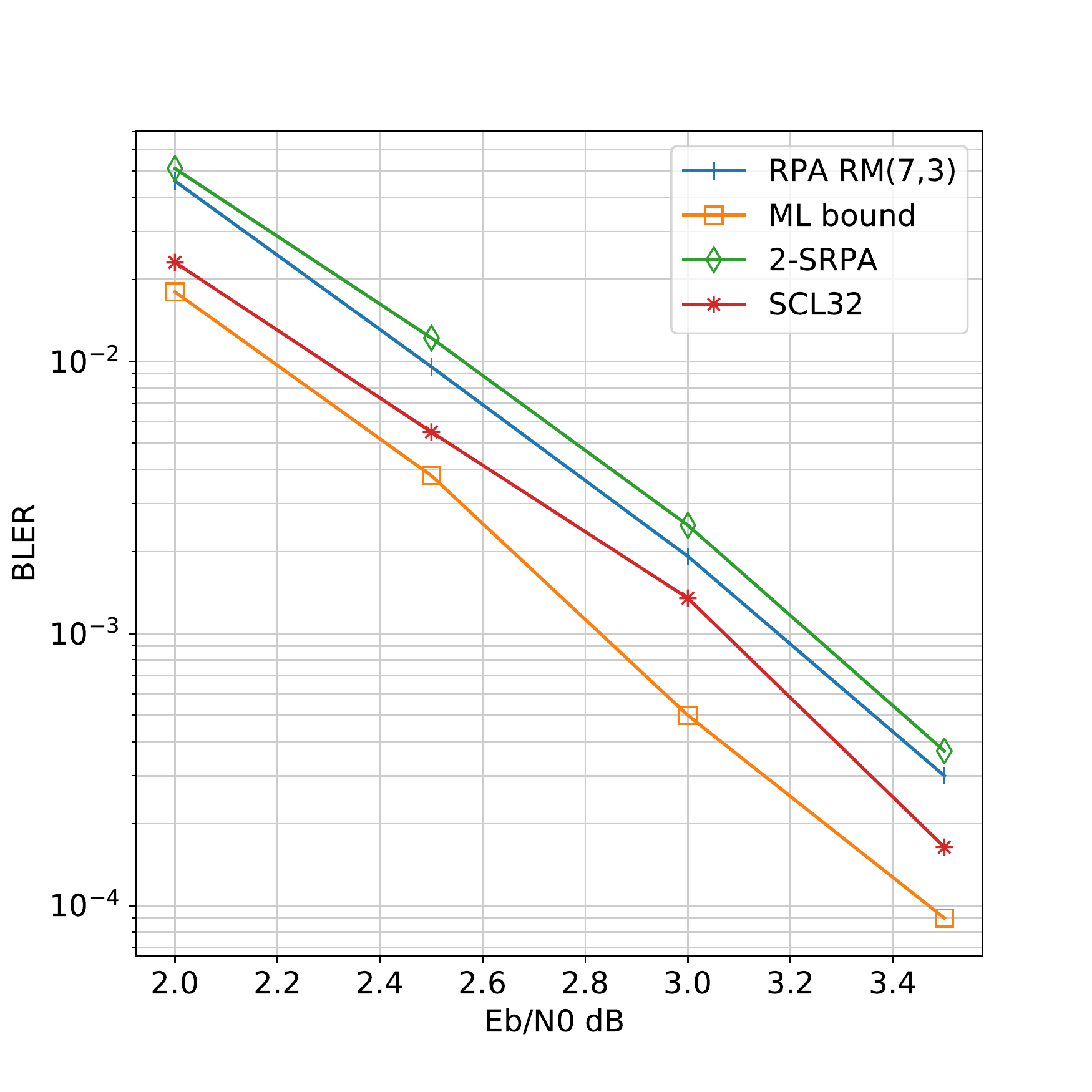}}  
\subfloat[$\mathcal{R} \mathcal{M} ( 8, 3 ) $]{\includegraphics[width=2.3in]{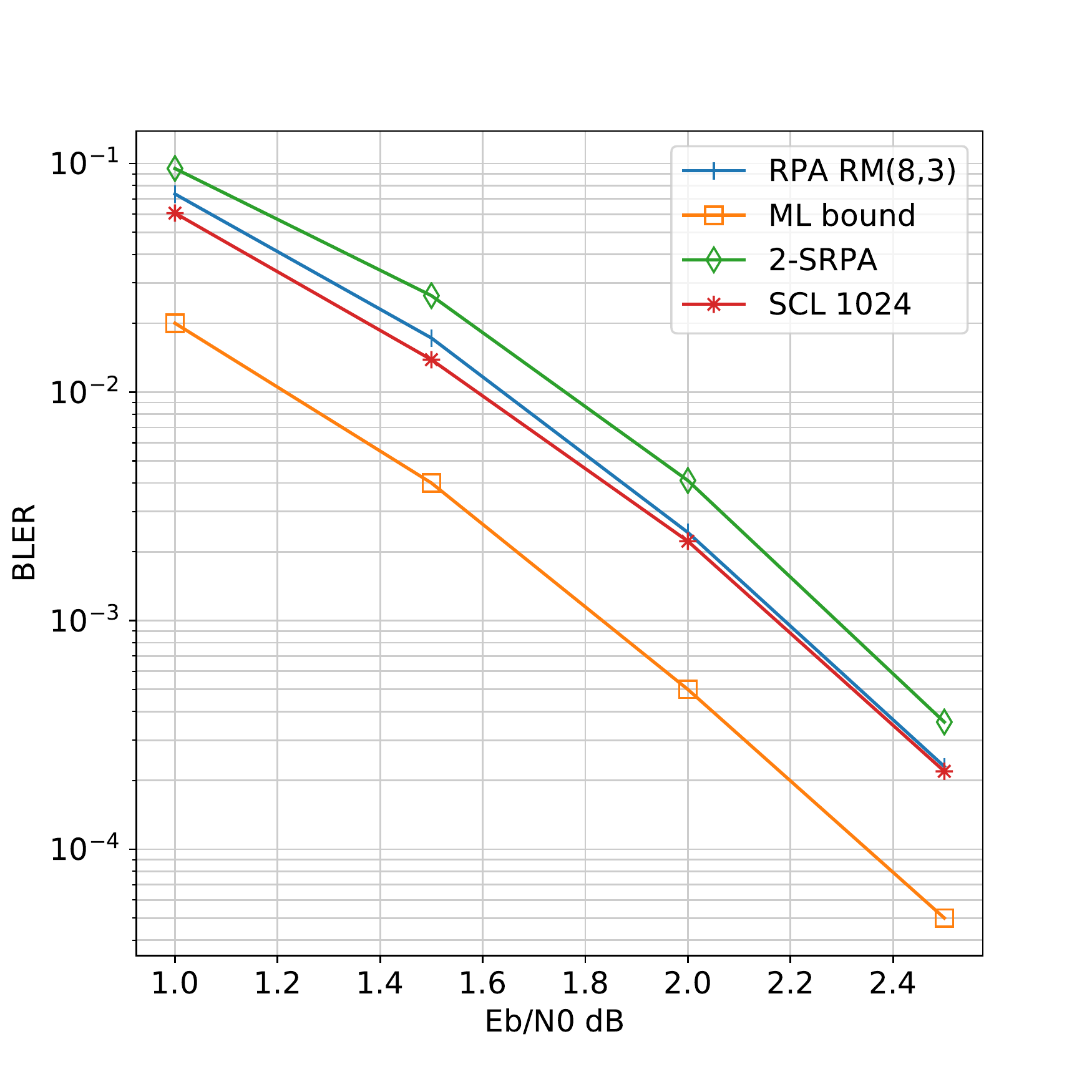}} 
\caption{ Performance comparison among $2$-SRPA decoding, RPA decoding, and SCL decoding. For $\mathcal{R} \mathcal{M} ( 8 , 2 ) $ and $\mathcal{R} \mathcal{M} ( 9 , 2 ) $, SRPA yields a performance improvement while requiring the same computational budget; for third-order RM codes, SCL performs better than RPA methods.}
\label{fig:3}
\end{figure*}
    
    Table~\ref{tab:1} provides a detailed comparison of the number of FHTs used by $2$-SRPA and RPA. We report \emph{(i)} the average number of FHTs used for RPA decoding at different $E_b/N_0$ levels where the early stopping condition is the same as \cite{RPAAY}, \emph{(ii)} the number of FHTs used by RPA when there is no early stopping (see the column \emph{Full Rounds}), and \emph{(iii)} the number of FHTs required for SRPA decoding. As SRPA decoding does not necessarily converge to a fixed point, it runs for a fixed number of iterations and does not require a comparison at different $E_b/N_0$ levels. The last column of Table~\ref{tab:1} provides the savings in the number of FHTs guaranteed by $2$-SRPA when compared to the RPA decoder with no early stopping (\emph{Full Rounds}). It can be seen that the proposed SRPA decoder can save up to $87\%$ of the FHT operations with respect to RPA decoding. When comparing with the average number of FHT operations in RPA decoding at different $E_b/N_0$ levels, the $2$-SRPA decoder leads to a saving between $50\%$ and $79\%$ of RPA's computational budget, while having little to no performance loss.



Fig.~\ref{fig:3}  compares the performance of $2$-SRPA decoding with SCL decoding. Recall that, given an RM code of order $r$, decoding each of its first-order RM sub-codes takes $2^{-(r-1)}n \log (2^{-(r-1)}n )$ operations. Furthermore, SCL decoding has complexity $L n \log n$, where $L$ is the list size. Thus, in our comparison, we assume that the decoding complexity of SCL with list size $L$ is roughly the same as SRPA using $2^{r-1}L$ FHTs. For second-order RM codes, the list size for each code is chosen such that either the computational budget of SCL is roughly the same as that of SRPA, or SCL reaches the ML decoding bound with less computation. It is visible that the performance of SCL decoding deteriorates quickly as the block length increases: for $N=128$ and $N=256$, SCL slightly improves upon SPRA using the same computational budget; However, for $N=512$, SRPA provides a significant improvement.  This fact does not hold for third-order RM codes. As seen in plots for $\mathcal{R} \mathcal{M}(7,3)$ and $\mathcal{R} \mathcal{M}(8,3)$, the SCL decoder with list size $32$ and $1024$, respectively, outperforms RPA methods, while requiring a much smaller computational budget.

\section{Conclusions} \label{sec:conc}
RPA decoding incorporates the algebraic structure of RM codes by using projections. By sparsifying the decoder on each iteration, this paper suggests that there exist more efficient ways to utilize these projections to lower the computational power required. Future work includes determining whether there exists a strategy for choosing the recursions that minimizes the error probability.


\section*{Acknowledgments}
We thank Emmanuel Abb\'e and Min Ye for providing us the implementation of RPA decoding.
D.~Fathollahi and M.~Mondelli are partially supported by the 2019 Lopez-Loreta Prize. N.~Farsad is supported by Discovery Grant from the Natural Sciences and Engineering Research Council of Canada (NSERC) and Canada Foundation for Innovation (CFI), John R. Evans Leader Fund. S.~A.~Hashemi is supported by a Postdoctoral Fellowship from NSERC.

\bibliographystyle{IEEEtran}
\bibliography{references}
\end{document}